\documentclass[pra,twocolumn,10pt]{revtex4-1}
\usepackage{graphicx}
\usepackage{dcolumn}
\usepackage{color}
\usepackage{times}
\usepackage{bm}
\usepackage{amssymb}
\usepackage{amsmath}
\usepackage{epsfig}
\usepackage{epstopdf}
\usepackage{dsfont}
\begin{document}
	\renewcommand{\thefootnote}{\fnsymbol {footnote}}
	
	\title{\textbf{Necessary and sufficient criterion of steering for two-qubit T states}}
	
	\author{Xiao-Gang Fan}
	\affiliation{School of Physics and Material Science, Anhui University, Hefei 230601, China}
	
	\author{Huan Yang}
	\affiliation{School of Physics and Material Science, Anhui University, Hefei 230601, China}
	
	\author{Fei Ming}
	\affiliation{School of Physics and Material Science, Anhui University, Hefei 230601, China}
	
	\author{Xue-Ke Song}
	\affiliation{School of Physics and Material Science, Anhui University, Hefei 230601, China}
	
	\author{Dong Wang}
	\affiliation{School of Physics and Material Science, Anhui University, Hefei 230601, China}
	
	\author{Liu Ye}
	\email{yeliu@ahu.edu.cn}
	\affiliation{School of Physics and Material Science, Anhui University, Hefei 230601, China}
	
\begin{abstract}
Einstein-Podolsky-Rosen (EPR) steering is the ability that an observer persuades a distant observer to share entanglement by making local measurements. Determining a quantum state is steerable or unsteerable remains an open problem. Here, we derive a new steering inequality with infinite measurements corresponding to an arbitrary two-qubit T state, from consideration of EPR steering inequalities with $N$ projective measurement settings for each side. In fact, the steering inequality is also a sufficient criterion for guaranteering that the T state is unsteerable. Hence, the steering inequality can be viewed as a necessary and sufficient criterion to distinguish whether the T state is steerable or unsteerable. In order to reveal the fact that the set composed of steerable states is the strict subset of the set made up of entangled states, we prove theoretically that all separable T states can not violate the steering inequality. Moreover, we
put forward a method to estimate the maximum violation from concurrence for arbitrary two-qubit T states, which indicates that the T state is steerable if its concurrence exceeds $1/4$.
\end{abstract}
\maketitle

\section{Introduction}
Schr\"odinger initially introduced the concept of steering in 1935 \cite{S1}, which is a significant  nonclassical phenomenon. And it formalized what Einstein called “spooky action at a distance" \cite{Einstein}. Although, Einstein-Podolsky-Rosen (EPR) paradox \cite{C1,C2,C3} was explored for a long time, the investigations concering steering has
been received extensive attentions till recently \cite{C9}. As one of nonclassical correlations, steering sits between entanglement and Bell nonlocality \cite{C4,C5,C6,C7}, and the intrinsical asymmetry is a nontrivial characteristic of steering, which is different from entanglement and Bell nonlocality \cite{C8,C9}. The preliminary works indicate that steering has a number of practical applications, such as one-sided device independent quantum key distribution \cite{C10,C11}, subchannel discrimination \cite{C12},  various protocols in quantum information processing \cite{C29}, and so on.

With the development of examinations about steering, a variety of sufficient criteria for detecting steering have been derived \cite{C13,C14,C15,C16,C17,C18,C19,C20,S4,S6,S5,B2,yang}. As long as one of these steering inequalities is violated, it can be used as a criterion for entanglement witness. All unsteerable states are Bell local, since a local hidden state (LHS) model \cite{C4,C23} is a particular case of a local hidden variable model. Historically, Wiseman \textit{et al.}  demonstrated that Werner state with weak entanglement does not violate any steering inequalities \cite{C4}. Whereafter, based on finite number of bilateral measurement experiments, Saunders \textit{et al.} experimentally proved that some steerable states are Bell local \cite{S3}. Besides, Bowles \textit{et al.} put forward a sufficient criterion for guaranteeing that a two-qubit state is unsteerable \cite{C22}. Nguyen \textit{et al.} show that quantum steering can be viewed as an inclusion problem in convex geometry \cite{C23}.  However, most of these results obtain sufficient criteria for steering or unsteering, and many criteria are only applicable to given numbers of measurement settings and outcomes.

In this paper, we put forward a new steering inequality with infinite measurements corresponding to an arbitrary two-qubit T state. And the steering inequality is a necessary and sufficient criterion that is used to make sure whether an arbitrary two-qubit T state is steerable or unsteerable.  To test the correctness of this steering inequality, we prove theoretically that all separable T states must follow it. In addition,  we establish the function relation between the concurrence and maximum violation for some special T states, and put forward a method to estimate the maximum violation from concurrence for any two-qubit T states.

\section{Preliminaries}
\subsection{Conditional states and LHS models}
Consider two distant observers Alice and Bob, who share a bipartite quantum state $\rho$ with reduced states $\rho_{\rm A} $ and $\rho_{\rm B}$. It is supposed that Alice can perform different measurements ${M_{\bm{r}}}=\bm{r}\cdot {\bm{\sigma}}$ on her assigned system. Here $\bm{r}$ serves as an arbitrary measurement setting (or measurement direction), and ${\bm{\sigma}}=\left( {{\sigma _x},{\sigma _y},{\sigma _z}} \right)$ is a matrix-vector consisting of three Pauli matrices. To be general, each such measurement ${M_{\bm{r}}}$ is described by the set of operators ${\left\{ {{M_{{\bm{r}}|a}}} \right\}_a}$ with the outcome $a$.  For each of  Alice's measurement setting $\bm{r}$ and outcome $a$, Bob retains a unnormalized conditional state ${\sigma _{\bm{r}|a}} = {\text{T}}{{\text{r}}_{\text{A}}}\left[ {\left( {{M_{\bm{r}|a}} \otimes \mathds{1}_2} \right)\rho } \right]$ with the probability $p\left( {\bm{r}|a} \right)={\text{Tr}}\left( {{\sigma _{\bm{r}|a}}} \right)$, where $\mathds{1}_2$ is the unit matrix of rank-2. And the conditional state obey the condition $\sum\nolimits_a {{\sigma _{\bm{r}|a}}}  = {\rho _{\text{B}}}$.

However, Bob is sceptical that Alice can remotely steer his state. And he is unsure whether he has received half of an entangled pair or a pure state sent by Alice. In order to eliminate this doubt, Bob tests whether the conditional state ${\sigma _{\bm{r}|a}}$ conforms to a LHS model. In other words, if the state ${\sigma _{\bm{r}|a}}$ obey the LHS model, there exists a probability density distribution function $p\left( {\bm{r}|a,\bm{v}} \right)$, which makes that the state ${\sigma _{\bm{r}|a}}$ can be expressed as \cite{C4,C22,C23}
\begin{align}
{\sigma _{\bm{r}|a}} =\frac{1}{{4\pi }} \iint\limits_S  {p\left( {\bm{r}|a,\bm{v} } \right){\sigma_{\bm{v}}}dS}, \label{01}
\end{align}
where the distribution function $p\left( {\bm{r}|a,\bm{v}} \right)$ is parametrized by the unit Bloch vector $\bm{v}=\left( {\begin{array}{*{20}{c}}{\sin \theta \cos \varphi },&{\sin \theta \sin \varphi },&{\cos \theta } \end{array}} \right)$, measurement setting $\bm{r}$ and outcome $a$. Here $dS=\sin \theta d \theta d \varphi$ represents the surface element, and the local hidden state $\sigma_{\bm{v}}$ denotes a normalized state that related to the Bloch vector $\bm{v}$. It is obvious that the probability $p\left( {\bm{r}|a} \right)$ can be represented by the integral of the probability distribution function $p\left( {\bm{r}|a,\bm{v}} \right)$, i.e.,
\begin{align}
p\left( {\bm{r}|a} \right) =\frac{1}{{4\pi }} \iint\limits_S  {p\left( {\bm{r}|a,\bm{v} } \right)dS}. \label{02}
\end{align}
If a representation as in Eq. \eqref{01} exists, Bod does not need to assume any kind of action at a distance to explain the post-measurement states ${\rho _{\bm{r}|a}}={\sigma_{\bm{r}|a}}/p\left( {\bm{r}|a} \right)$. Consequently, Alice fails to convince Bob that she can steer his system by her measurements, and one also says that the state $\rho$ is unsteerable from Alice to Bob. If such a model does not exist, Bob is required to believe that Alice can steer the state in his laboratory by some action at a distance. In this case, the state is said to be steerable from Alice to Bob.

\subsection{Steering inequalities with $N$ measurements}
For qubits, Alice's and Bob's $k$th measurement settings correspond to the measurement  ${M_{\bm{r}_k}}=\bm{r}_{k}\cdot \bm{\sigma}$ and ${M_{\bm{s}_k}}=\bm{s}_{k}\cdot \bm{\sigma}$, respectively. Here the measurement settings $\bm{r}_{k}$, $\bm{s}_{k}$ are unit vectors with three-dimension. The steering inequalities \cite{S3} can be expressed as
\begin{align}
{F_N}\left( \rho  \right) = \frac{1}{N}\sum\limits_{k = 1}^N {A_k}{\left\langle {{M_{\bm{s}_k}}} \right\rangle }  \leqslant {C_N}, \label{03}
\end{align}
where the random variable ${A_k} \in \left\{ { - 1,1} \right\}$ represents Alice's corresponding declared result for all $k$, and $\left\langle {M_{\bm{s}_k}} \right\rangle $ is the expected value of measurement ${M_{\bm{s}_k}}$ in the normalized conditional state ${\rho _{\bm{r}_{k}|a_k}}={\sigma_{\bm{r}_{k}|a_k}}/p\left( {\bm{r}_k|a_k} \right)$. For the sake of description, we call the quantity $F_N\left( \rho  \right)$ as the steering parameter for $N$ measurement settings. The bound $C_N$ is the maximum value $F_N$ can have if Bob has a pre-existing state known to Alice. And this bound can be denoted as
\begin{align}
{C_N} = \mathop {\max }\limits_{\left\{ {{A_k}} \right\}}  {{\lambda _{\max }}\left(O_N \right)}, \label{04}
\end{align}
where ${\lambda _{\max }}\left(  O_N  \right)$ stands for the largest eigenvalue of operator $O_N= {\frac{1}{N}\sum\limits_{k = 1}^N {{A_k}{M_{\bm{s}_k}}} } $.

If a two-qubit state $\rho$ violates the steering inequality in Eq. \eqref{03}, then the state $\rho$  must be steerable from Alice to Bob. However, if a two-qubit state $\rho$ conforms the steering inequality, then the state $\rho$ may be steerable or unsteerable from Alice to Bob.  In fact, Saunders \textit{et al.} \cite{S3} gave some bounds $C_N$, such as $C_2=1/\sqrt 2 $, $C_3=C_4=1/\sqrt 3 $, ${C_6} \approx 0.5393$, ${C_{10}} \approx 0.5236$ and so on. And their result shows that it should be possible to demonstrate steering if $\alpha>C_N$, for Werner states ${W_\alpha } = \alpha \left| {{\varphi _{\rm{B}}}} \right\rangle \langle {\varphi _{\rm{B}}}| + \frac{{1 - \alpha }}{4}{\mathds{1}_4 }$. Here the state $\left| {{\varphi _{\rm{B}}}} \right\rangle $ is one of Bell states, $\mathds{1}_4$ denotes the unit matrix with rank-4, and the parameter $\alpha$ represents the probability of $\left| {{\varphi _{\rm{B}}}} \right\rangle $. It indicates that the steering inequality can detect more and more steerable states, when the number $N$ of measurements increases.

\section{Results}
\subsection{Steering inequality with infinite measurements}
\begin{figure}
	\centering
	\includegraphics[width=8.2cm]{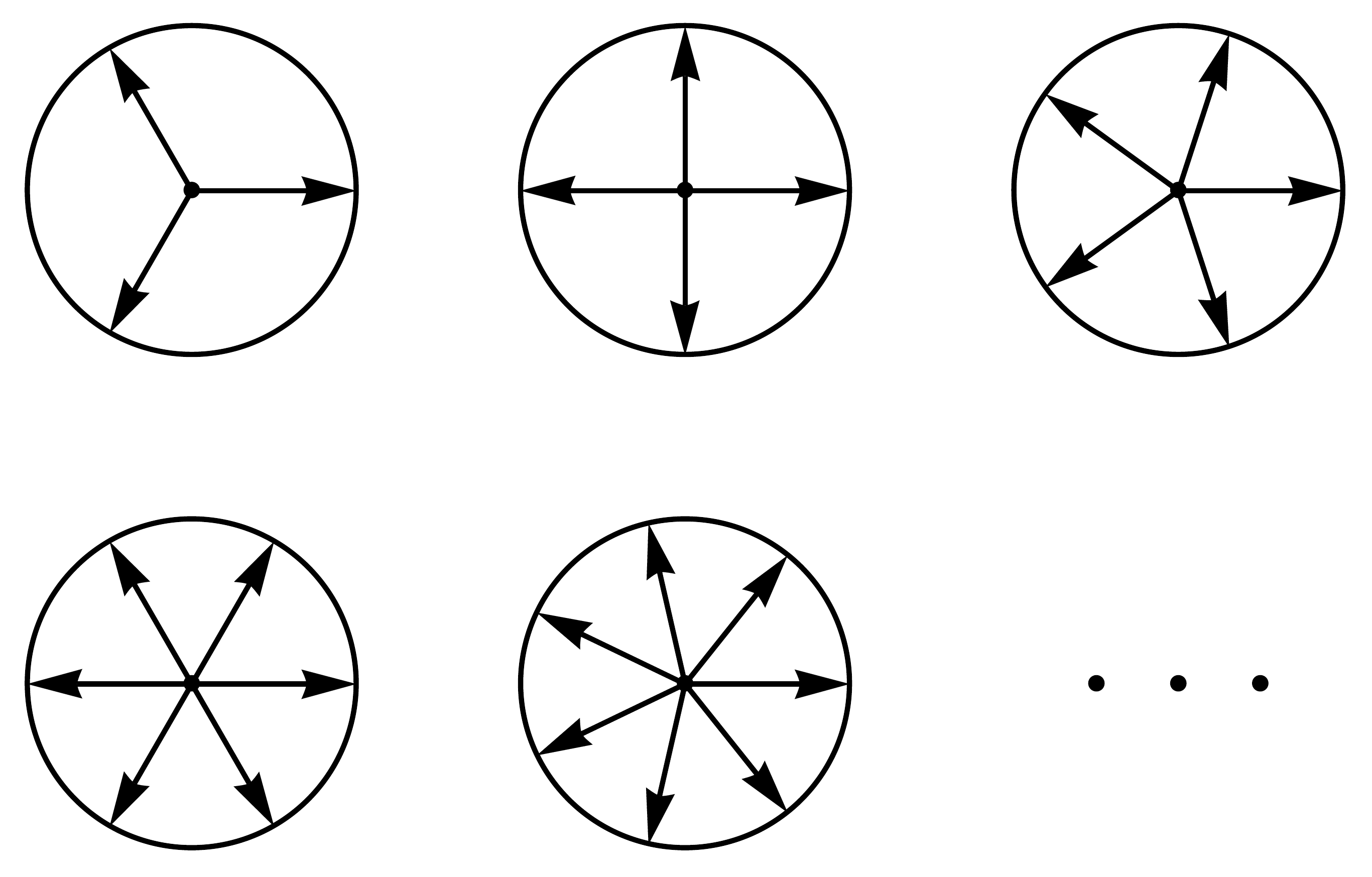}
	\caption{A round plate which is divided into $N$ equal parts by $N$ vectors. When it expands to the unit sphere and $N \to \infty $, we can consider $N$ unit vectors as $N$ cones with an infinitesimal bottom area $dS$. If the sum of the volumes about $N$ cones is equal to $4\pi /3$, then the sum of the bottom areas about $N$ cones is approximately equal to $4\pi$. Hence, when $N \to \infty $, $N$ vectors can divide the whole spherical surface into $N$ surface elements $dS$.}
	\label{Fig1}
\end{figure}
In order to improve the above inequalities in Eq. \eqref{03}, we consider a limiting case, i.e., $N \to \infty $. Clearly, two key issues need to be addressed. The first question is how to acquire this bound $C = \mathop {\operatorname{limit} }\limits_{N \to \infty } {C_N}$. And the second question is how to get the maximum violation of the steering inequality in the case of $N \to \infty $, i.e, how to obtain a limit maximum violation $F\left( \rho  \right) = \mathop {\operatorname{limit} }\limits_{N \to \infty } \left[ {\mathop {\max }\limits_{ {{\bm{r}_k}} } {F_N}\left( \rho  \right)} \right]$ of the steering inequality in Eq. \eqref{03}. For short, we call $F\left( \rho  \right)$  as the maximum violation.  Obviously, the maximum violation $F\left( \rho  \right)$ denotes the  maximum of the steering parameter $\mathop {\operatorname{limit} }\limits_{N \to \infty } {F_N}\left( \rho  \right)$ with infinite measurements. Just to keep the following derivation simple and easy to understand, we can write the same vector in two different ways: $\bm{u} = \left( {{u_1},{u_2},{u_3}} \right)$ and $\left| \bm{u} \right\rangle  = {\left( {\begin{array}{*{20}{c}}{{u_1}}&{{u_2}}&{{u_3}} \end{array}} \right)^{\text{T}}}$, where the superscript symbol T represents transpose of a matrix. 

On the one hand, we need to calculate this bound  $C = \mathop {\operatorname{limit} }\limits_{N \to \infty } {C_N}$. And the bound  $C$ is based on infinite measurement settings. In order to derive the limit bound $C$, we suppose that $N$ unit vectors $\left|{{\bm{s}_k}} \right\rangle $, which start at the sphere centre and end at the sphere surface, are uniformly distributed in Bloch sphere. When the number $N$ tends to infinity, these $N$ vectors can divide the whole spherical surface into $N$ surface elements $dS$ (as shown in Fig. 1).   For the sake of description, we replace $\left|{{\bm{s}_k}} \right\rangle $ with a variable unit vector $\left| \bm{v} \right\rangle  = {\left( {\begin{array}{*{20}{c}}{\sin \theta \cos \varphi }&{\sin \theta \sin \varphi }&{\cos \theta }\end{array}} \right)^{\text{T}}}$. And then the surface element is denoted as $dS = \sin \theta d\theta d\varphi \approx  4\pi /N$.

Based on the bound ${C_N}$, we consider the value of random variable $A_k$ in the following way to maximize the bound $C_N$. The way is: when the angle between the measurement setting $\bm{s}_k$ and the z-axis is acute, we set the random variable $A_k=1$; when the angle between the measurement setting $\bm{s}_k$ and the z-axis is obtuse, we set the random variable $A_k=-1$. Hence, if $N \to \infty $, the operator $O=\mathop {{\mathop{\rm limit}\nolimits} }\limits_{N \to \infty }O_N$ can be rewritten as
\begin{align}
O &= \mathop {{\mathop{\rm limit}\nolimits} }\limits_{N \to \infty } \frac{1}{N}\sum\limits_{k = 1}^N {{A_k}{M_{\bm{s}_k}}} \nonumber \\&=\frac{1}{{4\pi }}\left( {\iint\limits_{{S_\uparrow}} {{M_{\bm{v}}}dS} + \iint\limits_{{S_\downarrow}} { - {M_{\bm{v}}}dS}} \right), \label{05}
\end{align}
where $M_{\bm{v}}=\bm{v}\cdot \bm{\sigma}$ and $\bm{v}=\left( {\sin \theta \cos \varphi ,\sin \theta \sin\varphi ,\cos\theta } \right)$. Here $S_\uparrow$ (or $S_\downarrow$) stands for the upper (or lower) half Bloch spherical surface. It is easy to see that the operator $O$ in Eq. \eqref{05} can be reduced to $O={\sigma _z}/2$. Hence, the bound $C$ can be reduced to
\begin{align}
C={\lambda _{\max }}\left(  O  \right)=\frac{1}{2}. \label{06}
\end{align}

On the other hand, we need to obtain the  maximum violation $F\left( \rho  \right) = \mathop {\operatorname{limit} }\limits_{N \to \infty } \left[ {\mathop {\max }\limits_{ {{\bm{r}_k}} } {F_N}\left( \rho  \right)} \right]$. However, for a general two-qubit state $\rho$,  the  maximum violation $F\left( \rho  \right) $ is hard to be obtained. Therefore, in the following discussion, we will only consider the steering criterion of two-qubit T states. In general, based on Pauli operators $\bm{\sigma}=\left( {{\sigma _x},{\sigma _y},{\sigma _z}} \right)$, an arbitrary two-qubit T state $\rho$ can be expressed as
\begin{align}
\rho  = \frac{1}{4} ( \mathds{1}_4 + \sum\limits_{m,n } {{T_{mn}}{\sigma _m} \otimes {\sigma _n}}), \label{07}
\end{align}
where all information of the state $\rho$ is encoded into the elements $T_{mn}={\rm{Tr}}\left[ {\rho \left( {{\sigma _m} \otimes {\sigma _n}} \right)} \right]$ of correlation matrix $T\left( \rho  \right) $. Here, $m$ and $n \in \left\{ {x,y,z} \right\}$. It is obvious that the expected value $\left\langle {M_{\bm{s}_k}} \right\rangle $ can be denoted as
\begin{align}
\left\langle {M_{\bm{s}_k}} \right\rangle ={\text{Tr}}\left( {{\rho _{\bm{r}_{k}|a_k}} {M_{\bm{s}_k}}} \right)= \left\langle {{{{\overset{\lower0.5em\hbox{$\smash{\scriptscriptstyle\frown}$}}{\bm{b}} }_k}}}\mathrel{\left | {\vphantom {{{{\overset{\lower0.5em\hbox{$\smash{\scriptscriptstyle\frown}$}}{b} }_k}} {{s_k}}}}\right. \kern-\nulldelimiterspace}{{{\bm{s}_k}}} \right\rangle, \label{08}
\end{align}
where ${\overset{\lower0.5em\hbox{$\smash{\scriptscriptstyle\frown}$}}{\bm{b}} _k} = {\text{Tr}}\left( {{\rho _{\bm{r}_{k}|a_k}}\bm{\sigma}} \right)$ is Bloch vector of the conditional state ${\rho _{\bm{r}_{k}|a_k}}$. And the expected value $\left\langle {M_{\bm{s}_k}} \right\rangle $  can be given by \cite{C24}
 \begin{align}
 \left\langle {M_{\bm{s}_k}} \right\rangle= {a_k}\left\langle {{\bm{r}_k}}\right|T\left( \rho  \right) \left| {{\bm{s}_k}} \right\rangle. \label{09}
 \end{align}
  In order to be simple and not affect the final result, we might as well take $A_k=a_k$, which means that Alice declare the outcomes of her own measurements. Hence, the steering parameter in Eq. \eqref{03}  can be simplified as
\begin{align}
{F_N}\left( \rho  \right) = \frac{1}{N}\sum\limits_{k = 1}^N {\left\langle {{\bm{r}_k}} \right|T\left( \rho  \right) \left| {{\bm{s}_k}} \right\rangle }  . \label{10}
\end{align}

Obviously, when the unit vector  $\left| {{\bm{r}_k}} \right\rangle $  is collinear with the applied vector ${T\left( \rho  \right) }\left| {{\bm{s}_k}} \right\rangle $,  the value ${\left\langle {{\bm{r}_k}} \right|T\left( \rho  \right) \left| {{\bm{s}_k}} \right\rangle }$ in Eq. \eqref{10} can be maximized to be $X_k\left( \rho  \right)  = \mathop {\max }\limits_{\left| {{\bm{r}_k}} \right\rangle } \left\langle {{\bm{r}_k}} \right|T\left( \rho  \right) \left| {{\bm{s}_k}} \right\rangle $. Just for the sake of description, we set that the collinear condition of two vectors $ \left| {{\bm{r}_k}} \right\rangle $ and ${T\left( \rho  \right) }\left| {{\bm{s}_k}} \right\rangle $ is  written as $ u_k\left( \rho  \right) \left| {{\bm{r}_k}} \right\rangle = {T\left( \rho  \right) }\left| {{\bm{s}_k}} \right\rangle $. Here, $u_k\left( \rho  \right) >0$ stands for the collinear coefficient. Based on the property $\left\langle {{{\bm{r}_k}}}\mathrel{\left | {\vphantom {{{r_k}} {{r_k}}}}\right. \kern-\nulldelimiterspace}{{{\bm{r}_k}}} \right\rangle  = 1$, the collinear coefficient $u_k\left( \rho  \right) $ is denoted as $u_k\left( \rho  \right)  =\sqrt{\left\langle {{\bm{s}_k}} \right|{T^{\text{T}}\left( \rho  \right) }T\left( \rho  \right) \left| {{\bm{s}_k}} \right\rangle }$. Hence, the maximum expected value $X_k\left( \rho  \right) $ of $k$th measurement outcome for an arbitrary T state $\rho$ can be indicated as $X_k\left( \rho  \right)  =u_k\left( \rho  \right) =\sqrt{\left\langle {{\bm{s}_k}} \right|{T^{\text{T}}\left( \rho  \right) }T\left( \rho  \right) \left| {{\bm{s}_k}} \right\rangle }$. Based on the maximum expected value $X_k\left( \rho  \right) $, the maximum of the steering parameter in Eq. \eqref{10} can be rewritten as
\begin{align}
\mathop {\max }\limits_{\left| {{\bm{r}_k}} \right\rangle } {F_N}\left( \rho  \right)  = \frac{1}{N}\sum\limits_{k = 1}^N { \sqrt {\left\langle {{\bm{s}_k}} \right|{T^{\text{T}}\left( \rho  \right) }T\left( \rho  \right) \left| {{\bm{s}_k}} \right\rangle }}. \label{11}
\end{align}
In the same way, we replace $\left|{{\bm{s}_k}} \right\rangle $ with a variable unit vector $\left| \bm{v} \right\rangle  = {\left( {\begin{array}{*{20}{c}}{\sin \theta \cos \varphi }&{\sin \theta \sin \varphi }&{\cos \theta }\end{array}} \right)^{\text{T}}}$. Consequently, the  maximum violation $F\left( \rho  \right) = \mathop {\operatorname{limit} }\limits_{N \to \infty } \left[ {\mathop {\max }\limits_{ {{\bm{r}_k}} } {F_N}\left( \rho  \right)} \right]$ can be reduced to
\begin{align}
F\left( \rho  \right)=\frac{1}{{4\pi }}\iint\limits_S  {\sqrt {\left\langle \bm{v} \right|{T^{\text{T}}\left( \rho  \right) }T\left( \rho  \right) \left| \bm{v} \right\rangle } dS }, \label{12}
\end{align}
where $\bm{v}=\left( {\begin{array}{*{20}{c}}{\sin \theta \cos \varphi },&{\sin \theta \sin \varphi },&{\cos \theta } \end{array}} \right)$ and  $dS=\sin \theta d \theta d \varphi$. Combining Eqs.  \eqref{06} and \eqref{12}, we can obtain a steering inequality that can be described by the following \textit{Theorem 1}.

\textit{Theorem.} For the T state $\rho$, the steering inequality with infinite measurements can be expressed as
\begin{align}
\frac{1}{{2\pi }}\iint\limits_S  {\sqrt {\left\langle \bm{v} \right|{T^{\text{T}}\left( \rho  \right) }T\left( \rho  \right) \left| \bm{v} \right\rangle } dS } \le 1. \label{13}
\end{align}

\textit{Theorem} shows that the T state $\rho$ is steerable if $F\left( \rho  \right)>1/2$. And it is based on infinite measurements corresponding to an arbitrary two-qubit T state. Therefore, it is the best optimization of steering inequalities in Eq. \eqref{03} with finite number of measurements. In other words, it can detect more steerable states than Eq. \eqref{03}.  Specially, Werner state $W_\alpha$ is steerable if the probability $\alpha$ conforms to the relation $\alpha  > 1/2$.

\subsection{Properties of the  maximum violation}

According to the calculation result of the  maximum violation $F\left( \rho  \right)$, we now provide two important properties that are scaling and symmetry.

\textit{Property 1.} Given a two-qubit T state $\rho$, we consider a family of states ${\rho _\gamma }$ by mixing it with a special kind of separable noise,
\begin{align}
{\rho _\gamma } = \gamma \rho +\left( {1 - \gamma } \right)\frac{\mathds{1}_4 }{4},
\end{align}
where $0 \leqslant \gamma  \leqslant 1$. For these states ${\rho _\gamma }$, we can show that
\begin{align}
F\left( {\rho _\gamma }  \right)=\gamma F\left( \rho \right). \label{15}
\end{align}

\textit{Proof.} In combination with the  calculation formula $T_{mn}={\rm{Tr}}\left[ {\rho \left( {{\sigma _m} \otimes {\sigma _n}} \right)} \right]$, the relation between the correlation matrixs $T\left( \rho  \right) $ and $T\left( {\rho _\gamma }  \right) $ can be given by $T\left( {\rho _\gamma }  \right)=\gamma T\left( \rho \right)$. Thus, the derivative process can be given by
\begin{align}
F\left( {\rho _\gamma }  \right)&=\frac{1}{{4\pi }}\iint\limits_S  {\sqrt {\left\langle \bm{v} \right|{T^{\text{T}}\left( {\rho _\gamma }  \right) }T\left( {\rho _\gamma }  \right) \left| \bm{v} \right\rangle } dS }\nonumber \\&=\frac{\gamma}{{4\pi }}\iint\limits_S  {\sqrt {\left\langle \bm{v} \right|{T^{\text{T}}\left( \rho  \right) }T\left( \rho  \right) \left| \bm{v} \right\rangle } dS }\nonumber \\&=\gamma F\left( \rho \right).
\end{align}

\textit{Property 2.} Given a two-qubit T state $\rho$, we consider a family of states $\rho '$ which are formed by the unitary operation applied to the original state $\rho$,
\begin{align}
\rho ' = \left( {{U_{\text{A}}} \otimes {U_{\text{B}}}} \right)\rho \left( {U_{\text{A}}  \otimes U_{\text{B}}} \right)^\dag,
\end{align}
where $U_\text{A}$ and $U_\text{B}$ are the unitary matrices on Alice's and Bob's side, respectively. For these states, we can show that
\begin{align}
F\left( \rho' \right)=F\left( \rho  \right). \label{18}
\end{align}

\textit{Proof.} When a local unitary operation  ${{U_{\text{A}}} \otimes {U_{\text{B}}}}$ is performed on a T state $\rho$, the final state $\rho'$ is also a T state \cite{C25,C26}. And the relation between the correlation matrixs $T\left( \rho  \right) $ and $T\left( \rho'  \right) $ can be given by  $T\left( {\rho '} \right) = {R_{\rm{A}}}T\left( \rho  \right)R_{\rm{B}}^{\rm{T}}$, where the elements of ${R_{\rm{A}}}$ and ${R_{\rm{B}}}$ can be denoted as ${\left( {{R_{\rm{A}}}} \right)_{kk'}} = \frac{1}{2}{\rm{Tr}}\left( {{\sigma _k}{U_{\rm{A}}}{\sigma _{k'}}U_{\rm{A}}^\dag } \right)$ and ${\left( {{R_{\rm{B}}}} \right)_{kk'}} = \frac{1}{2}{\rm{Tr}}\left( {{\sigma _k}{U_{\rm{B}}}{\sigma _{k'}}U_{\rm{B}}^\dag } \right)$, respectively. And ${R_{\rm{A}}}$ and ${R_{\rm{B}}}$ belong to the three-dimensional rotation group $SO\left( 3 \right)$. We rewrite the  maximum violation of the final state as
\begin{align}
F\left( \rho'  \right)&=\frac{1}{{4\pi }}\iint\limits_S  {\sqrt {\left\langle \bm{v} \right|{T^{\text{T}}\left( \rho'  \right) }T\left( \rho'  \right) \left| \bm{v} \right\rangle } dS } \nonumber \\&=\frac{1}{{4\pi }}\iint\limits_S  {\sqrt {\left\langle \bm{v} \right|{R_{\rm{B}}}{T^{\text{T}}\left( \rho  \right) }T\left( \rho  \right)R_{\rm{B}}^{\rm{T}} \left| \bm{v} \right\rangle } dS }\nonumber \\&=\frac{1}{{4\pi }}\iint\limits_S  {\sqrt {\left\langle \bm{v'} \right|{T^{\text{T}}\left( \rho  \right) }T\left( \rho  \right) \left| \bm{v'} \right\rangle } dS },
\end{align}
where $\left| \bm{v'} \right\rangle=R_{\rm{B}}^{\rm{T}} \left| \bm{v} \right\rangle $ denotes a new unit vector. To represent the suface element $dS'$ corresponding to the new vector $\left| \bm{v'} \right\rangle$, we set $\left| \bm{v'} \right\rangle = {\left( {\begin{array}{*{20}{c}}{\sin \theta' \cos \varphi' }&{\sin \theta' \sin \varphi' }&{\cos \theta' }\end{array}} \right)^{\text{T}}}$. Obviously $dS'=\sin \theta' d \theta' d \varphi'$. Notice that the new suface element $dS'$ is given by the original suface element $dS$ by the rotation operation $R_{\rm{B}}^{\rm{T}}$, which indicates $dS'=dS$. Therefore, local unitary operation does not change the  maximum violation. The derivative process can be described as follows
\begin{align}
F\left( \rho'  \right)&=\frac{1}{{4\pi }}\iint\limits_S  {\sqrt {\left\langle \bm{v'} \right|{T^{\text{T}}\left( \rho  \right) }T\left( \rho  \right) \left| \bm{v'} \right\rangle } dS }\nonumber \\&=\frac{1}{{4\pi }}\iint\limits_S  {\sqrt {\left\langle \bm{v'} \right|{T^{\text{T}}\left( \rho  \right) }T\left( \rho  \right) \left| \bm{v'} \right\rangle } dS' }\nonumber \\&=\frac{1}{{4\pi }}\iint\limits_S  {\sqrt {\left\langle \bm{v} \right|{T^{\text{T}}\left( \rho  \right) }T\left( \rho  \right) \left| \bm{v} \right\rangle } dS }\nonumber \\&=F\left( \rho \right).
\end{align}
Here, the derivation takes advantage of the the property that the integral is independent of the integral variable.

\textit{Inference.} For an arbitrary two-qubit T state $\rho$, the  maximum violation $F\left( \rho  \right)$ is only related to three singular values $\left\{ {{t_1}\left( \rho  \right),{t_2}\left( \rho  \right),{t_3}\left( \rho  \right)} \right\}$ of the correlation matrix $T\left( \rho  \right) $. And the formula can be expressed as follows
\begin{align}
F\left( \rho  \right) = \frac{1}{{4\pi }}\iint\limits_S {\sqrt {\left\langle \bm{v} \right|{\Lambda ^2}\left( \rho  \right)\left| \bm{v} \right\rangle } dS}, \label{21}
\end{align}
where ${\Lambda}\left( \rho  \right)=diag\left\{ {{t_1}\left( \rho  \right),{t_2}\left( \rho  \right),{t_3}\left( \rho  \right)} \right\}$ represents a diagonal matrix consisting of these singular values.

\textit{Proof.} For a general two-qubit T state $\rho$, there is a local unitary operation that transforms the state $\rho$ into a Bell diagonal state $\rho_{\rm{Bell}}$, whose correlation matrix $T\left( \rho_{\rm{Bell}}  \right) $ satisfies the relation ${T^2}\left( {{\rho _{{\rm{Bell}}}}} \right) = {\Lambda ^2}\left( \rho  \right)$. Therefore, combining with \textit{Property 2}, we obtain 
\begin{align}
F\left( \rho  \right) &= F\left( \rho_{\rm{Bell}}  \right)=\frac{1}{{4\pi }}\iint\limits_S {\sqrt {\left\langle \bm{v} \right|{T^2}\left( \rho_{\rm{Bell}}  \right)\left| \bm{v} \right\rangle } dS}\nonumber \\&=\frac{1}{{4\pi }}\iint\limits_S {\sqrt {\left\langle \bm{v} \right|{\Lambda ^2}\left( \rho  \right)\left| \bm{v} \right\rangle } dS}.
\end{align}

\subsection{Sufficient criterion for unsteerablity}
For each of  Alice's measurement setting $\bm{r}$ and outcome $a$, Bob retains a conditional state ${\rho_{\bm{r}|a}}={\sigma_{\bm{r}|a}}/p\left( {\bm{r}|a} \right)$. And the eigenvalues of ${\rho_{\bm{r}|a}}$ can be reduced to
\begin{align}
{\lambda _1}&= \frac{{1 + \sqrt {\left\langle \bm{r} \right|T\left( \rho  \right){T^{\rm{T}}}\left( \rho  \right)\left| \bm{r} \right\rangle } }}{2},
\nonumber \\
{\lambda _2} &= \frac{{1 - \sqrt {\left\langle \bm{r} \right|T\left( \rho  \right){T^{\rm{T}}}\left( \rho  \right)\left| \bm{r} \right\rangle } }}{2}.
\end{align}
In order to acquire the sufficient criterion of unsteering for an arbitrary T state, we start with a any Bell diagonal state $\rho_{\rm{Bell}}$. 

According to `\textit{proof of Theorem 1}' in Ref. \cite{C22}, we conclude that if the conditional state of $\rho_{\rm{Bell}}$ conforms to a LHS model, then its eigenvalues ${\lambda _1}$ and ${\lambda _2}$ satisfy the relation ${\lambda _1}\le2\sqrt{{\lambda _2}}-{\lambda _2}$ for the measurement setting $\bm{r}$, or equivalently
\begin{align}
\sqrt {\left\langle \bm{r} \right|{T^2}\left( {{\rho _{{\rm{Bell}}}}} \right)\left| \bm{r} \right\rangle }  \le \frac{1}{2}. \label{a}
\end{align}
Therefore, when we consider infinite ($N \to \infty $) mesurements, Eq. \eqref{a} can be rewritten as
\begin{align}
\mathop {{\mathop{\rm limit}\nolimits} }\limits_{N \to \infty } \frac{1}{N}\sum\limits_{k = 1}^N {\sqrt {\langle {\bm{r}_k}|{T^2}\left( {{\rho _{{\rm{Bell}}}}} \right)\left| {{\bm{r}_k}} \right\rangle } }  \le \frac{1}{2}. \label{b}
\end{align}
Similarly, we replace $\left|{{\bm{r}_k}} \right\rangle $ with a variable unit vector $\left| \bm{v} \right\rangle  = {\left( {\begin{array}{*{20}{c}}{\sin \theta \cos \varphi }&{\sin \theta \sin \varphi }&{\cos \theta }\end{array}} \right)^{\text{T}}}$. And then, Eq. \eqref{b} can be reduced to 
\begin{align}
\frac{1}{{4\pi }}\iint\limits_S {\sqrt {\left\langle \bm{v} \right|{T^2}\left( \rho_{\rm{Bell}}  \right)\left| \bm{v} \right\rangle } dS}\le \frac{1}{2}.
\end{align}
Thus, it is unsteerable for Bell diagonal state $\rho_{\rm{Bell}}$, if $F\left( \rho_{\rm{Bell}}  \right) \le 1/2$. In fact, this result is consistent with the criteria given in Ref. \cite{C24}.

Considering the symmetry (\textit{Property 2} and its \textit{Inference}) of the  maximum violation $F\left( \rho  \right)$, we obtain a sufficient criterion that it is unsteerable for a general T state $\rho$ if $F\left( \rho \right) \le 1/2$. And \textit{Theorem} shows that the T state $\rho$ is steerable if $F\left( \rho  \right)>1/2$. Therefore, the T state $\rho$ is steerable if only and if the  maximum violation meets the relation $F\left( \rho  \right) >1/2$.

\subsection{Separable states don't violate the steering inequality}

In general, the steerable states must be entangled. In other words, the separable states must be unsteerable. Obviously, there is a rule that the separable states must obey the steering inequality. Therefore, it is necessary to test the newly derived steering inequality in Eq. \eqref{13}. 

\textit{Rule.} If the T state $\rho$ is a separable state, then it conforms the steering inequality in Eq. \eqref{13}.

\textit{Proof.} In general, a separable T state can be expressed as ${\rho} = \sum\nolimits_i {{p_i}{\rho _i}}$ with Bloch vectors $\bm{a}={\rm{Tr}}\left[ {\rho \left( {\bm{\sigma}  \otimes \mathds{1}_2 } \right)} \right]=\bm{0}$ and $\bm{b}={\rm{Tr}}\left[ {\rho \left( {\mathds{1}_2 \otimes \bm{\sigma}  } \right)} \right]=\bm{0}$,  where ${\rho _i} = \rho _i^{\text{A}} \otimes \rho _i^{\text{B}}$ denotes $i$th uncorrelated state and the probability $p_i$ satisfies the relation $\sum\nolimits_i {{p_i}}  = 1$. In orther words, these vectors  $\bm{a}_i={\rm{Tr}}\left( {\rho_{i}^{\text{A}} {\bm{\sigma}} } \right)$ and $\bm{b}_i={\rm{Tr}}\left( {\rho_{i}^{\text{B}}\bm{\sigma}} \right)$ meet the relations $\sum\nolimits_i {{p_i} {{\bm{a}_i}}} =\bm{0}$ and $\sum\nolimits_i {{p_i} {{\bm{b}_i}}} =\bm{0}$. And the correlation matrix $T\left( \rho  \right)$ can be described by these vectors, i.e., $T\left( \rho  \right)=\sum\nolimits_i {{p_i}\left| {{\bm{a}_i}} \right\rangle \left\langle {{\bm{b}_i}} \right|}$. Combining with the integrand $f\left( \rho,\bm{v} \right)=\sqrt {\left\langle \bm{v} \right|{T^{\text{T}}}\left( \rho  \right)T\left( \rho  \right)\left| \bm{v} \right\rangle }$, we obain that the integrand $f\left( \rho,\bm{v} \right)$ can be rewritten as
\begin{align}
f\left( \rho,\bm{v} \right) = \sqrt {\sum\limits_{i,j} {{p_i}{p_j}\left\langle {\bm{v}}\mathrel{\left | {\vphantom {v {{b_i}}}}\right. \kern-\nulldelimiterspace}{{{\bm{b}_i}}} \right\rangle \left\langle {{{\bm{a}_i}}}\mathrel{\left | {\vphantom {{{a_i}} {{a_j}}}}\right. \kern-\nulldelimiterspace}{{{\bm{a}_j}}} \right\rangle \left\langle {{{\bm{b}_j}}}\mathrel{\left | {\vphantom {{{b_j}} v}}\right. \kern-\nulldelimiterspace}{\bm{v}} \right\rangle } }. \label{23}
\end{align}
Considering the relation $\left| {\left\langle {{{\bm{a}_i}}}\mathrel{\left | {\vphantom {{{a_i}} {{a_j}}}}\right. \kern-\nulldelimiterspace}{{{\bm{a}_j}}} \right\rangle } \right| \leqslant 1$, we rewrite Eq. \eqref{23} as an inequality
\begin{align}
f\left( \rho,\bm{v} \right) &\leqslant \sqrt {\sum\limits_{i,j} {{p_i}{p_j}\left| {\left\langle {\bm{v}}\mathrel{\left | {\vphantom {s {{b_i}}}}\right. \kern-\nulldelimiterspace}{{{\bm{b}_i}}} \right\rangle \left\langle {{{\bm{a}_i}}}\mathrel{\left | {\vphantom {{{a_i}} {{a_j}}}}\right. \kern-\nulldelimiterspace}{{{\bm{a}_j}}} \right\rangle \left\langle {{{\bm{b}_j}}}\mathrel{\left | {\vphantom {{{b_j}} s}}\right. \kern-\nulldelimiterspace}{\bm{v}} \right\rangle } \right|} } \nonumber \\&\leqslant \sqrt {\sum\limits_{i,j} {{p_i}{p_j}} \left| {\left\langle {\bm{v}}\mathrel{\left | {\vphantom {s {{b_i}}}}\right. \kern-\nulldelimiterspace}{{{\bm{b}_i}}} \right\rangle \left\langle {{{\bm{b}_j}}}\mathrel{\left | {\vphantom {{{b_j}} s}}\right. \kern-\nulldelimiterspace}{\bm{v}} \right\rangle } \right|}  \nonumber \\&=\sqrt {\sum\limits_i {{p_i}\left| {\left\langle {\bm{v}}\mathrel{\left | {\vphantom {s {{b_i}}}}\right. \kern-\nulldelimiterspace}{{{\bm{b}_i}}} \right\rangle } \right|} \sum\limits_j {{p_j}\left| {\left\langle {\bm{v}}\mathrel{\left | {\vphantom {s {{b_j}}}}\right. \kern-\nulldelimiterspace}{{{\bm{b}_j}}} \right\rangle } \right|} }  \nonumber \\&= \sum\limits_i {{p_i}\left| {\left\langle {\bm{v}}\mathrel{\left | {\vphantom {s {{b_i}}}}\right.\kern-\nulldelimiterspace}{{{\bm{b}_i}}} \right\rangle } \right|}.
\end{align}
It is obvious that the  maximum violation $F\left( \rho  \right)$ for separable T state $\rho$ satisfies the following relation
\begin{align}
F\left( \rho  \right)&=\frac{1}{{4\pi }}\iint\limits_S {f\left( \rho,\bm{v} \right)dS} \nonumber \\&\leqslant \frac{1}{{4\pi }}\iint\limits_S {\sum\limits_i {{p_i}\left| {\left\langle {\bm{v}}\mathrel{\left | {\vphantom {s {{b_i}}}}\right.\kern-\nulldelimiterspace}{{{\bm{b}_i}}} \right\rangle } \right|}dS}\nonumber \\&=\frac{1}{2}\sum\limits_i {{p_i}\left( {\frac{1}{{2\pi }}\iint\limits_S {\left| {\left\langle {\bm{v}}\mathrel{\left | {\vphantom {s {{b_i}}}}\right.\kern-\nulldelimiterspace}{{{\bm{b}_i}}} \right\rangle } \right|dS}} \right)}\nonumber \\&=\frac{1}{2}\sum\limits_i {{p_i}\left( {\frac{1}{{2\pi }}\iint\limits_S {\left| {{b_i}} \right|\left| {\cos \theta } \right|dS}} \right)}\nonumber \\&=\frac{1}{2}\sum\limits_i {{p_i}\left| {{b_i}} \right|}\int_0^\pi  {\left| {\cos \theta } \right|\sin \theta d\theta }  \nonumber \\&=\frac{1}{2}\sum\limits_i {{p_i}\left| {{b_i}} \right|}\le \frac{1}{2}\sum\limits_i {{p_i}}=\frac{1}{2}. 
\end{align}
Therefore, all separable T states don't violate the steering inequality in Eq. \eqref{13}.

\subsection{Relation between the concurrence and maximum violation}

We now illustrate the boundary problem of the intrinsic relation between steering and entanglement with some special T states, and try to estimating the maximum violation by using entanglement. In order to better understand the relation between two quantum correlations, we introduce a common measure of entanglement for two-qubit states, i.e., concurrence \cite{C27}. For an arbitrary pure state $\left| \psi  \right\rangle $, its concurrence can be defined as
\begin{align}
E\left( {\left| \psi  \right\rangle } \right) = \left| {\left\langle {\psi }\mathrel{\left | {\vphantom {\psi  {\tilde \psi }}}\right. \kern-\nulldelimiterspace}{{\tilde \psi }} \right\rangle } \right|, \label{26}
\end{align}
where $\left| {\tilde \psi } \right\rangle {\rm{ = (}}{\sigma _y} \otimes {\sigma _y})\left| {{\psi ^*}} \right\rangle $ represents the spin-flipped state of $\left| \psi  \right\rangle $ and $\left| {{\psi ^*}} \right\rangle $ is the complex conjugate state of $\left| \psi  \right\rangle $. For a general T state $\rho $, its spin-flipped state $\tilde \rho  = \left( {{\sigma _y} \otimes {\sigma _y}} \right){\rho ^*}\left( {{\sigma _y} \otimes {\sigma _y}} \right)$ is same as the state $\rho$, i.e., $\tilde \rho  =\rho$. Thus, the concurrence of $\rho$ can be expressed as \cite{C25,C27,C28}
\begin{align}
E\left( \rho  \right) &=\max \left\{ {0,2{\lambda _{\max }}\left( {\sqrt {\sqrt \rho  \tilde \rho \sqrt \rho  } } \right) - {\rm{Tr}}\left( {\sqrt {\sqrt \rho  \tilde \rho \sqrt \rho  } } \right)} \right\}
\nonumber \\&= \max \left\{ {0,2{\lambda _{\max }}\left( \rho  \right) - 1} \right\}, \label{27}
\end{align}
where ${\lambda _{\max }}\left(  X  \right)$ represents the maximum eigenvalue of the matrix $X$.

(1) \textit{Evolutionary states of Werner states.}--- We consider the evolutionary states ${W_{\rm{PD}}}$, which are formed by Werner states ${W_\alpha}$ going through the phase damped (PD) channel. And the states ${W_{\rm{PD}}}$ can be denoted as
\begin{align}
{W_{\rm{PD}}}=\sum\limits_{i = 0}^1 {{K_i}{W_\alpha}K_i^\dag  }, \label{28}
\end{align}
where ${K_0} = \left| 0 \right\rangle \left\langle 0 \right|{\rm{ + }}\sqrt {1 - \eta } \left| 1 \right\rangle \left\langle 1 \right|$ and ${K_1} =\sqrt \eta  \left| 1 \right\rangle \left\langle 1 \right|$ are the Kraus operators of PD channel. Obviously, the states ${W_{\rm{PD}}}$ belong to  Bell diagonal states. Based on Eq. \eqref{27}, the concurrence can be expressed as $E\left( {{W_{{\text{PD}}}}} \right) = \max \left\{ {0,\alpha \sqrt {1 - \eta }  - \frac{{1 - \alpha }}{2}} \right\}$.  And the correlation matrix of ${W_{\rm{PD}}}$ can be written as a diagonal matrix, i.e., $T\left( {{W_{{\text{PD}}}}} \right) = diag\left\{ {\begin{array}{*{20}{c}}{\alpha \sqrt {1 - \eta } },&{ - \alpha \sqrt {1 - \eta } },&\alpha \end{array}} \right\}$. Hence, the maximum violation $F\left( {W_{\rm{PD}}}  \right)$ can be reduced to
\begin{align}
F\left( {{W_{{\text{PD}}}}} \right) = \frac{\alpha }{2}\left( {1 + \frac{{1 - \eta }}{{\sqrt \eta  }}\ln \frac{{1 + \sqrt \eta  }}{{\sqrt {1 - \eta } }}} \right), \label{29}
\end{align}
where $\ln $ denotes the nature-logarithm. Thus, the sufficient and necessary criterion of steering for the states ${W_{\rm{PD}}}$ is the relation $ \alpha \left( {1 + \frac{{1 - \eta }}{{\sqrt \eta  }}\ln \frac{{1 + \sqrt \eta  }}{{\sqrt {1 - \eta } }}} \right)>1$. In particular, when $\eta  = 0$, the states ${W_{\rm{PD}}}$ are reduced to Werner states ${W_\alpha}$. And the maximum violation corresponding to Werner states ${W_\alpha}$ can be given by $F\left( {{W_\alpha }} \right) = \mathop {\operatorname{limit} }\limits_{\eta  \to 0} F\left( {{W_{{\text{PD}}}}} \right) = \alpha $. It is apparent that Werner states $W_\alpha$ are to demonstrate steering if and only if $\alpha>1/2$ (as shown in Fig. 2).

\begin{figure}
	\centering
	\includegraphics[width=8.2cm]{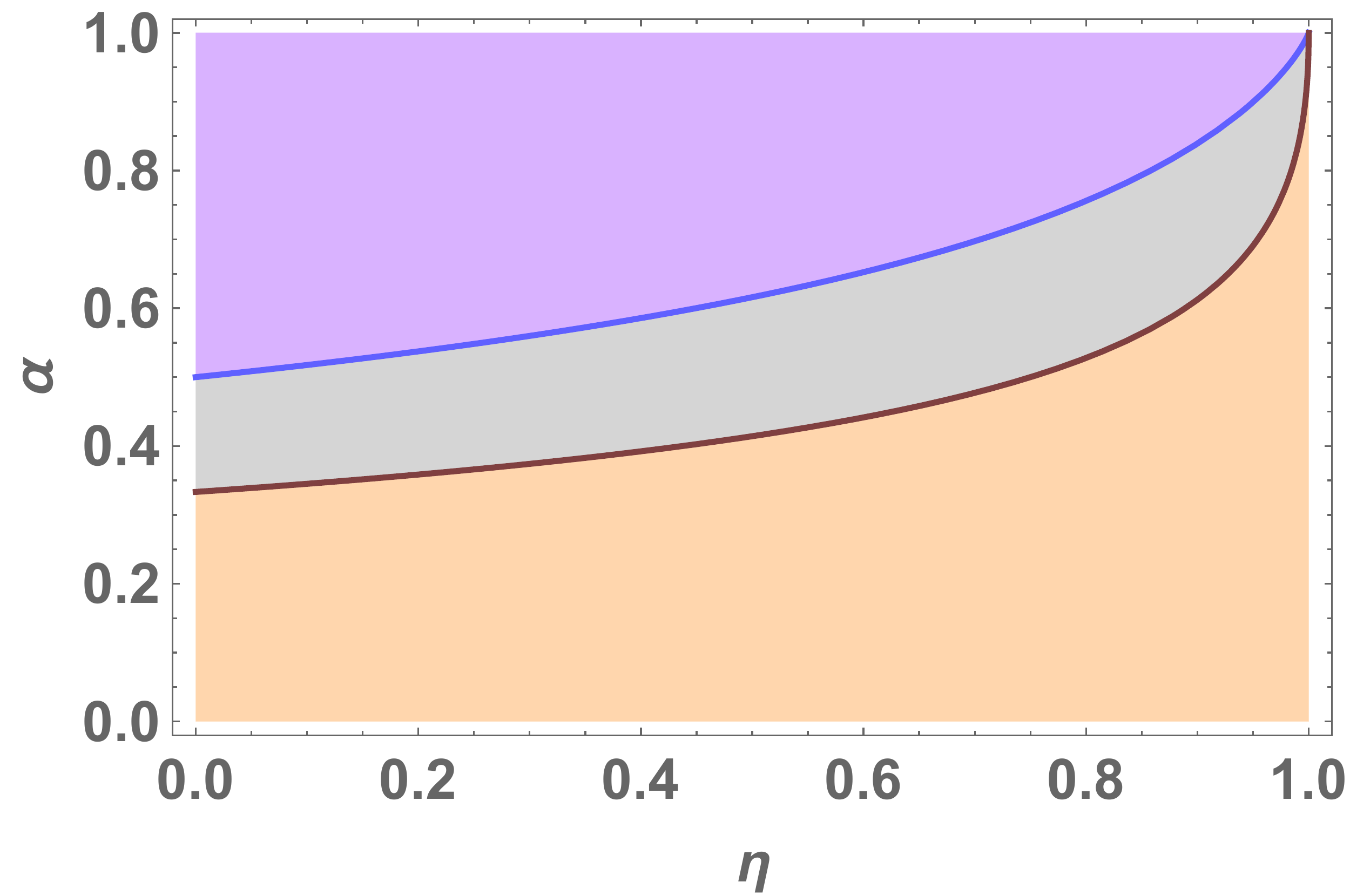}
	\caption{(Color online) Characterization of concurrence and steering for the states ${W_{\rm{PD}}}$. The brown solid line represents the boundary between the entangled and separable states. And the blue solid line denotes the boundary between the steerable and unsteerable states, obtained from Eq. \eqref{29}. The light orange region represents the set of separable states. The states are entangled but unsteerable in the gray region. And the states are steerable in the light purple region.}
	\label{Fig1}
\end{figure}

(2)\textit{ T states of rank-2.}--- Based on the lemma that an arbitrary two-qubit state has a decomposition in which each pure state has the same entanglememt \cite{C25}, an arbitrary two-qubit T state ${\psi _{\text{T}}}$ of rank-2 can be formed by mixing an arbitrary pure state $\left| \psi  \right\rangle $ and its spin-flipped state $\left| {\tilde \psi } \right\rangle$ with equal probability. And the T states ${\psi _{\text{T}}}$ can be written as
\begin{align}
{\psi _{\text{T}}} = \frac{1}{2}\left( {\left| \psi  \right\rangle \left\langle \psi  \right| + \left| {\tilde \psi } \right\rangle \left\langle {\tilde \psi } \right|} \right). \label{30}
\end{align}
According to Eqs. \eqref{26} and \eqref{27}, we obtain an invariability that concurrence of the constructed T state ${\psi _{\text{T}}}$ is equal to concurrence of the original state $\left| \psi  \right\rangle $, i.e., $E\left( {{\psi _{\text{T}}}} \right) = E\left( {\left| \psi  \right\rangle } \right)$. For the state ${\psi _{\text{T}}}$, three singular values of the correlation matrix $T\left( {\psi _{\text{T}}} \right)$ can be given by $\Lambda \left( {{\psi _{\text{T}}}} \right) = diag\left\{ {\begin{array}{*{20}{c}}1, &{E\left( {{\psi _{\text{T}}}} \right)}, &{E\left( {{\psi _{\text{T}}}} \right)} \end{array}} \right\}$. Therefore, the maximum violation for the state ${\psi _{\text{T}}}$ can be reduced to
\begin{align}
F\left( {{\psi _{\text{T}}}} \right) = \frac{1}{2}\left[ {1 + \frac{{{E^2\left( {{\psi _{\text{T}}}} \right)}}}{{\sqrt {1 - {E^2\left( {{\psi _{\text{T}}}} \right)}} }}\ln \frac{{1 + \sqrt {1 - {E^2\left( {{\psi _{\text{T}}}} \right)}} }}{E\left( {{\psi _{\text{T}}}} \right)}} \right]. \label{31}
\end{align}

In particular, when $E\left( {{\psi _{\text{T}}}} \right) = 0$, the maximum violation is $\mathop {\operatorname{limit} }\limits_{c  \to 0} F\left(\psi _{\text{T}} \right) = 1/2 $; when $E\left( {{\psi _{\text{T}}}} \right)  = 1$, the maximum violation is $\mathop {\operatorname{limit} }\limits_{c  \to 1} F\left(\psi _{\text{T}} \right) = 1 $. Hence, for an arbitrary T state with rank-2, the state is steerable if only and if this state is entangled.

(3)\textit{ Any T states.}--- At the front, we have given the fuction relation between the concurrence and maximum violation for some special states. For any two-qubit T states $\rho $, what relation could we obtain about the concurrence and maximum violation? When we only know the values of concurrence, where is the value-range of the maximum violation? In orther words, we try to establish an inequality relation between the concurrence and maximum violation, and use the concurrence to estimate the maximum violation. In order to accomplish this task, we investigate lots of randomly generated two-qubit T states. The result shows that there is an inequality relation between the concurrence and the maximum violation. For the separable T states, $0\leqslant F\left( \rho  \right)\leqslant1/2$. When $E\left( \rho  \right)>0$, the inequality can be expressed as follows (see Fig. 3)
\begin{widetext}
	\begin{align}
		\frac{{1 + 2E\left(\rho  \right)}}{3} \leqslant F\left(\rho  \right) \leqslant \frac{1}{2}\left[ {1 + \frac{{{E^2\left(\rho  \right)}}}{{\sqrt {1 - {E^2\left(\rho  \right)}} }}\ln \frac{{1 + \sqrt {1 - {E^2\left(\rho  \right)}} }}{E\left(\rho  \right)}} \right]. \label{32}
	\end{align}
\end{widetext}
Obviously, when $E\left(\rho  \right)>1/4$, the T state $\rho $ is steerable.

\begin{figure}
	\centering
	\includegraphics[width=8.2cm]{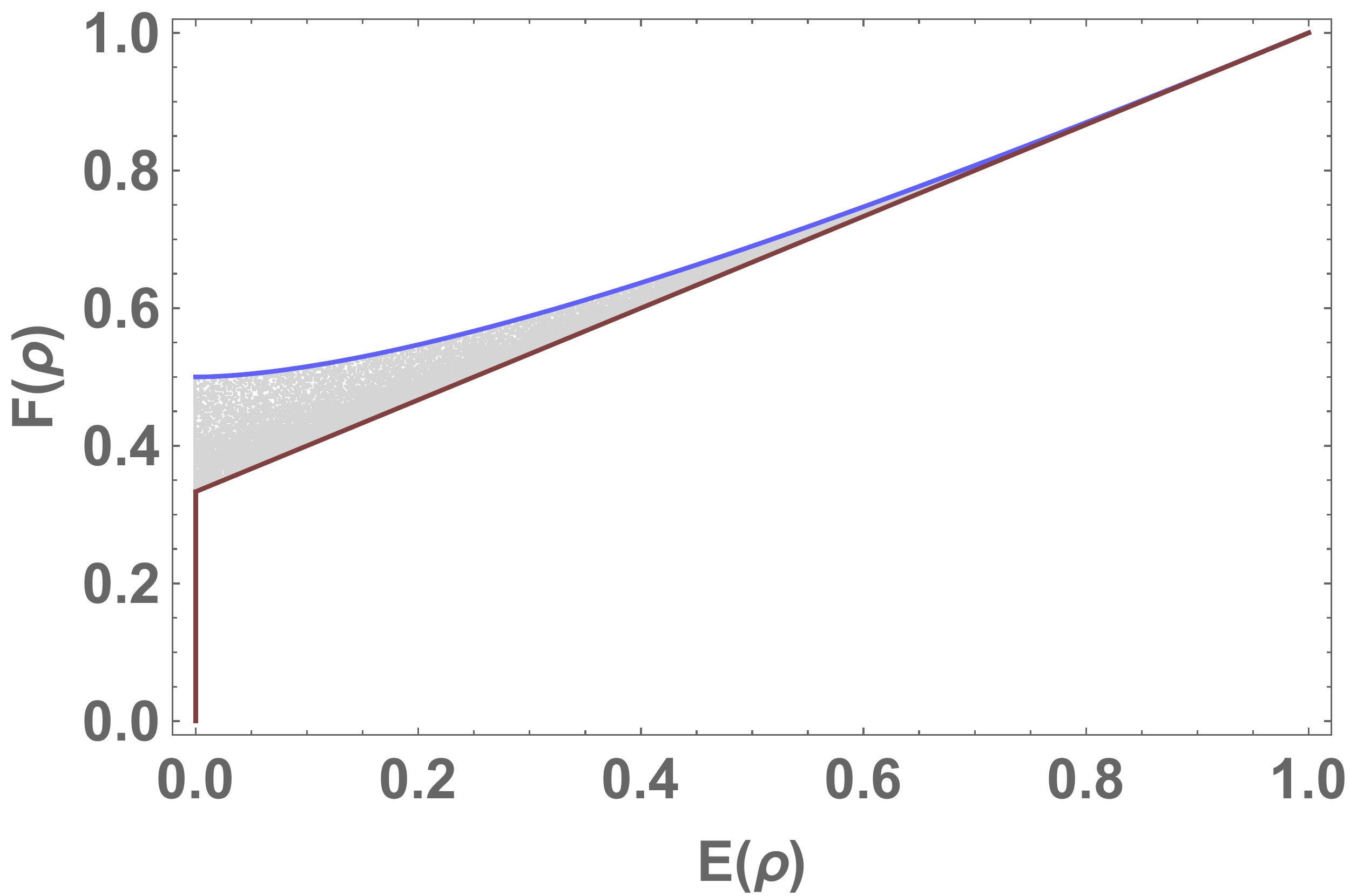}
	\caption{(Color online) Maximum violation  $F\left( \rho  \right)$ versus concurrence $E\left( \rho  \right)$ for two-qubit T states $\rho $. The upper bound (blue solid line) denotes Eq. \eqref{31}, which corresponds to the mixed states ${{\psi _{\text{T}}}}$. And the lower bound (brown solid line) denotes the relation $E\left( {{W_\alpha }} \right) = \max \left\{{\begin{array}{*{20}{c}}{0,}&{\frac{{3F\left( {{W_\alpha }} \right) -1}}{2}}\end{array}} \right\}$, which corresponds to Werner states $W_\alpha$. The figure plots the maximum violation  $F\left( \rho  \right)$, along the Y axis, and concurrence $E\left( \rho  \right)$, along the X axis, for $5 \times {10^4}$ randomly generated two-qubit T states, by using a specific Mathematica package.}
	\label{Fig1}
\end{figure}

\section{Conclusion and discussion}
In this paper, we have completed two main tasks about two-qubit T states. On the one hand, we derive a steering inequality with infinite measurements corresponding to an arbitrary two-qubit T state. And the steering inequality can be viewed as a necessary and sufficient criterion that is used to distinguish that the T state is steerable or unsteerable. And a two-qubit T state is steerable if and only if the maximum violation is beyond $1/2$. For an arbitrary two-qubit T state, the maximum violation satisfies the scaling and symmetry in Eqs. \eqref{15} and \eqref{18}. On the other hand, we establish the function relation between the concurrence and maximum violation for some special T states, and put forward a method to estimate the maximum violation from concurrence for any two-qubit T states by using lots of randomly generated two-qubit T states. And it indicates that an arbitrary T state is steerable if its concurrence exceeds $1/4$. Specially, for all T states with rank-2, the state is steerable if only and if this state is entangled.

In future work, it would be interesting to investigate the necessary and sufficient criterion that ensures an arbitrary two-qubit state is steerable or unsteerable from Alice to Bob.

\section*{Acknowledgements} 
This work was supported by the National Science Foundation of China under Grant No. 11575001.
\bibliographystyle{plain}

\end{document}